\documentclass[11pt,twoside]{article} 
\usepackage{asp2004}
\usepackage{epsf}
\usepackage{psfig}
\usepackage{lscape} 

\begin{document} 
\title{Kilo-gauss Magnetic Fields in three DA White Dwarfs}
\author{R. Aznar Cuadrado,$^1$ S. Jordan,$^2$ 
        R. Napiwotzki,$^3$ H. M. Schmid,$^4$ 
        S. K. Solanki,$^1$ and G. Mathys$^5$} 
\affil{$^1$Max-Planck-Institut f\"{u}r Sonnensystemforschung, 
       Max-Planck-Str. 2, 37191 Katlenburg-Lindau, Germany\\
       $^2$Astronomisches Rechen-Institut, M\"{o}nchhofstr. 12-14, 
       69120 Heidelberg, Germany\\
       $^3$Department of Physics \& Astronomy, University of Leicester, 
       University Road, Leicester LE1 7RH, UK\\
       $^4$Institut f\"{u}r Astronomie, ETH Zentrum, 
       8092 Z\"{u}rich, Switzerland\\
       $^5$European Southern Observatory, Casilla 19001, 
       Santiago 19, Chile\\}
\begin{abstract} 
We have detected longitudinal magnetic fields between 2 and 4\,kG in three 
normal DA white dwarfs (WD\,0446$-$790, WD\,1105$-$048, WD\,2359$-$434) out 
of a sample of 12 by using optical spectropolarimetry done with the VLT Antu 
8 m telescope equipped with FORS1. With the exception of 40 Eri B (4\,kG) 
these are the first positive detections of magnetic fields in white dwarfs 
below 30\,kG. A detection rate of 25 \% (3/12) may indicate now for the first 
time that a substantial fraction of white dwarfs have a weak magnetic field. 
This result, if confirmed by future observations, would form a cornerstone 
for our understanding of the evolution of stellar magnetic fields.
\end{abstract}

\section{Introduction}
On the main sequence and at later stages of evolution, magnetic fields have a 
major impact on the angular momentum loss and stellar winds, on building-up 
chemical anomalies and abundance inhomogeneities across the stellar surface, 
on convection and the related coronal activity, and other evolutionary 
processes especially in interacting binaries. The first detection of a 
magnetic field on a white dwarf (WD) was made by Kemp et al. (1970) on 
Grw+$70^\circ$ 8247, and large spectroscopic and polarimetric surveys have 
been carried out since then (e.g. Schmidt \& Smith 1995). The magnetic fields 
of WDs could simply be ``fossil'' remnants of the fields already present in 
main-sequence stars, but strongly amplified by contraction. This hypothesis 
assumes that the magnetic flux (e.g. through the magnetic equator) is 
conserved to a large extent during the stellar evolution.

In main-sequence stars magnetic fields have been detected directly mainly for 
peculiar magnetic Ap and Bp stars with rather well organized fields and field 
strengths $\sim$ $10^2-10^4$ G. For weak fields in A to O stars ($B<10^2$ G) 
direct magnetic field detections are still very rare. For sun-like stars 
ample evidence (coronal activity) for the presence of complicated small-scale 
fields exists, but direct measurements are only possible for the more active 
stars (e.g. Saar 1996; R\"{u}edi et al. 1997). The contraction to a WD 
amplifies the magnetic fields by about 4 orders of magnitude, so that weak 
and often undetectable magnetic fields on the main sequence become measurable 
during the WD phase. This is supported by the known magnetic WDs with 
Megagauss fields ($B=10^6-10^9$~G). Their frequency and spatial distribution, 
as well as their mass, are consistent with the widely accepted view that they 
are the descendents of the magnetic Ap and Bp stars (e.g. Mathys 2001). 
Magnetic main-sequence stars with weaker magnetic fields have been suggested 
as possible progenitor candidates for the magnetic degenerates with weaker 
fields (e.g. Schmidt et al. 2003). The B stars on which weaker fields have 
been detected may be the missing stars. However, even the most sensitive 
observations are limited to some tens of Gauss on main-sequence stars.

Thus, magnetic field amplification during stellar evolution may offer the 
opportunity to investigate $\sim 1$~G magnetic fields (averaged global 
fields) in normal main-sequence stars with observations of $\sim 1$~kG 
magnetic fields during the WD stage. Zeeman splitting of narrow NLTE 
line cores in the Balmer lines becomes undetectable in intensity spectra for 
weak fields ($<20$ kG) or for objects without narrow line core. Therefore, 
spectropolarimetry is the most promising technique for successful detections 
of weak magnetic fields. 

It is interesting to note that, recently, magnetic fields have also been 
detected in the direct progeny of white dwarfs, subdwarfs and central stars
of planetary nebulae (O'Toole et al. and Jordan et al., these proceedings).

\section{Observations and Data Reduction}
Spectropolarimetric observations of a sample of 12 normal DA WDs were carried 
out during the period 4 November 2002 -- 3 March 2003, in service mode by ESO 
staff members using FORS1 at the VLT. With a 0.8$\arcsec$ wide slit we 
obtained a (FWHM) spectral resolution of 4.5 \AA. The data were recorded 
using a Tek. 2048$\times$2048 CCD with 24 $\mu$m pixels which correspond to a 
pixel scale of 0.2$\arcsec$/pixel in spatial and 1 \AA/pixel in spectral 
direction. Spectra were acquired with grism G600B (spectral range 3400--6000 
\AA), covering all H\,{\sc i} Balmer lines from H$\beta$ to the Balmer jump 
simultaneously. 

Our 12  targets were selected with the criterion of not having any sign of 
Zeeman splitting visible in the SPY-UVES spectra (Napiwotzki et al. 2003), 
and hence no magnetic fields above a level of about 20\,kG. All our targets 
have strong hydrogen lines, ideal for measuring line polarisation, and no 
bright companion. Most of our targets were observed at least over two nights.

In all frames the bias level was subtracted and the frames were cleaned of 
cosmic ray hits. The frames were then flat-field corrected and the wavelength 
calibration was applied.

\section{Circular Polarisation}
In order to obtain circular polarisation spectra a Wollaston prism and a 
quarter-wave plate were inserted into the optical path. Each exposure yields 
two spectra on the detector, one from the extra-ordinary beam and the other 
from the ordinary beam. Stokes $V$ is obtained from a differential 
measurement of photon counts in either the ordinary or extra-ordinary beams, 
measured at two different angles of the retarder waveplate. We adopted the 
FORS1 standard observing sequence for circular polarimetry consisting of 
exposures with retarder plate position angles $+45^{\circ}$ and $-45^{\circ}$.

In order to derive the circular polarisation from a sequence of exposures, 
we added up the exposures with the same quarter-wave plate position angle. 
The Stokes ($V/I$) can be obtained as
\begin{equation}
\frac{V}{I} = \frac{(R-1)}{(R+1)}, ~~\mbox{ with }
R^2=\left(\frac{f_o}{f_e}\right)_{\alpha=+45}\times 
    \left(\frac{f_e}{f_o}\right)_{\alpha=-45}
\end{equation}
where $V$ is the Stokes parameter which describes the net circular 
polarisation, $I$ is the unpolarized intensity, $\alpha$ indicates the 
nominal value of the position angle of the retarder waveplate, and $f_o$ and 
$f_e$ are the fluxes on the detector from the ordinary and extra-ordinary 
beams of the Wollaston, respectively. 

\section{Determination of Weak Magnetic Fields}
For field strengths below 10\,kG the Zeeman splitting of the Balmer lines is 
less than 0.1 \AA. This is well below the width of the cores of the Balmer 
lines in all the stars of our sample (typically a few \AA). Therefore, we can 
apply the weak-field approximation (e.g., Landi degl'Innocenti \& Landi 
degl'Innocenti 1973) without any loss of accuracy. According to this 
approximation the measured $V$ and $I$ profiles are related to 
$\langle B_z\rangle$ by the expression:
\begin{equation}
\frac{V}{I} = -g_{\rm eff} \ensuremath{C_z}\lambda^{2}\frac{1}{I}
\frac{\partial I}{\partial \lambda}
\ensuremath{\langle\large B_z\large\rangle}\;,
\end{equation}
where $g_{\rm eff}$ is the effective Land\'{e} factor (= 1 for all hydrogen 
lines of any series, Casini \& Landi degl'Innocenti 1994), $\lambda$ is the 
wavelength expressed in \AA, $\langle B_z\rangle$ is the mean longitudinal 
component of the magnetic field  expressed in Gauss and the constant 
$C_z=e/(4\pi m_ec^2)$ $(\simeq 4.67 \times10^{-13}\,{\rm G}^{-1} \AA^{-1})$.
Note that this approximation also holds if instrumental broadening is 
present, but it is not generally correct if the profiles are rotationally 
broadened (Landstreet 1982).

The error associated with the determination of the longitudinal field 
obtained from individual Balmer lines is larger for Balmer lines at shorter 
wavelengths than for lines at longer wavelengths. This is due to the 
combination of two effects: while the Zeeman effect increases as lambda 
squared, most other line broadening effects depend linearly on lambda, so 
that the magnetic field is better detected at longer wavelengths than at 
shorter wavelengths; furthermore, the Balmer lines at shorter wavelengths are 
less deep, so that $\partial I/\partial \lambda$ is smaller. Using H$\beta$ 
and H$\gamma$ simultaneously, we obtained a determination of the mean 
longitudinal magnetic field that best fit the observed ($V/I$).

For stars observed during more than one night the circular polarisation 
spectra ($V/I$) flux weighted means were calculated. The averaging helps to 
extract signal hidden in the noise of the individual exposures if the stellar 
Zeeman signal remains unchanged with time. In two of the averaged spectra 
(corresponding to WD\,0446$-$789, see Fig. 1, and WD\,2359$-$434) a moderate 
S-shape circular polarisation signature across H$\beta$ and H$\gamma$ can be 
noticed. For WD\,1105$-$048 polarisation reversals of those lines were only 
present in one of the two observing nights. This may indicate a different 
orientation of the magnetic field between the two epochs due to stellar 
rotation.

To determine the longitudinal component of the magnetic field for each 
measurement we compared the observed circular polarisation for an interval 
of $\pm 20$\,\AA\ around H$\beta$ and H$\gamma$ with the prediction of 
Eq.\,2. The best fit for $\langle B_z\rangle$, the only free parameter, was 
found by a $\chi^2$-minimisation procedure. If we assume that no magnetic 
field is present, all deviations from zero polarization are due to noise. 
This can be expressed by the standard deviation $\sigma$ over the respective 
intervals around the Balmer lines. 
In Table\,1 the best fits for $B_z$ derived from the analysis of H$\beta$ 
and H$\gamma$ lines are shown. 
We find a significant magnetic field in WD\,2359$-$434, WD\,0446$-$789 (see 
Fig.\,1) and WD\,1105$-$048. For the first two stars the magnetic field is 
detected at the $3\sigma$ level individually from H$\beta$ and H$\gamma$, as 
well as from the combination of both lines. 
\begin{figure}[!t]
\epsfxsize=12.cm \epsfbox{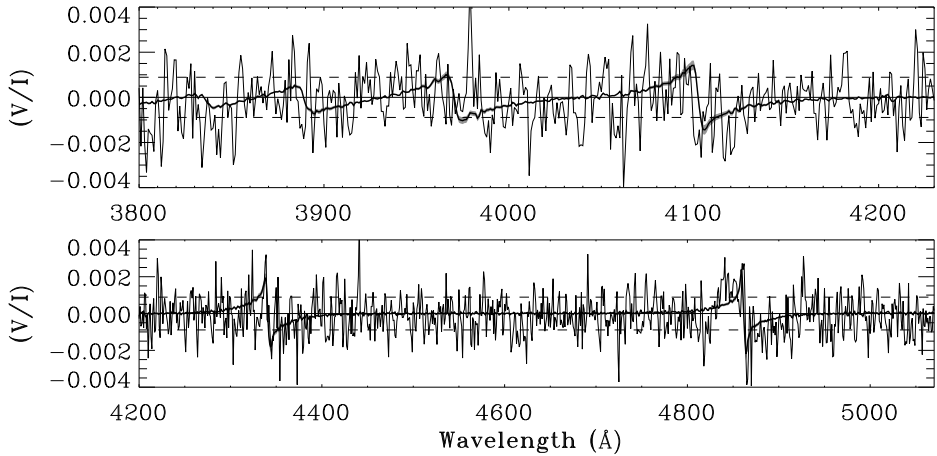}
\caption{($V/I$) spectra of WD\,0446$-$789 (thin solid line) in the region of 
H$\beta$ and H$\gamma$ (bottom) and close to the Balmer jump (top). The solid 
horizontal line indicates the zero level. The horizontal dashed lines 
indicate the position of the $1\sigma$ level of the average ($V/I$) spectrum. 
The light shading represents the variation between ($V/I$) spectra predicted 
by the low-field approximation (Eq.\,2) using $B$ values of 4283$\pm$640\,G 
(thick solid line).}
\end{figure}
Higher members of the Balmer series do not contain enough $V$-signal to give 
reliable results, but the analysis of H$\delta$ confirms our positive 
detections of magnetic fields.

\section{Atmospheric and Stellar Parameters}
Masses and cooling ages of WDs are of special interest to distinguish between 
the proposed formation scenarios. These quantities can be computed from the 
fundamental stellar parameters temperature and gravity, which can be derived 
by a model atmosphere analysis of the spectra, and theoretical cooling tracks.

The observed line profiles are fitted with theoretical spectra from a large 
grid of NLTE spectra calculated with the code developed by Werner (1986). 

The observed and theoretical Balmer line profiles are normalised to a linear 
continuum in a consistent manner. Radial velocity offsets are corrected by 
shifting the spectra to a common wavelength scale. The synthetic spectra are 
convolved to the resolution of the observed spectra (4.5\,\AA) with a 
Gaussian and interpolated to the actual parameters. The atmospheric 
parameters are then determined by minimising the $\chi^2$ value by means of 
a Levenberg-Marquardt steepest descent algorithm. This procedure is applied 
simultaneously to all Balmer lines of one observed spectrum. Errors are 
estimated to be 2.3\% in T$_{\rm eff}$ and 0.07\,dex in $\log g$ (Napiwotzki 
et al. 1999). 

The parameters listed in Table\,1 are the averages of the fit results for the 
individual spectra. WD masses were computed from a comparison of parameters 
derived from the fit with the grid of WD cooling sequences of Benvenuto \& 
Althaus (1999) for an envelope hydrogen mass of $10^{-4}M_{\mathrm{WD}}$. 
Spectroscopic distances $d$(spec) are determined from the absolute magnitudes 
computed for the given stellar parameters using the synthetic $V$ band fluxes 
computed by Bergeron et al. (1995) and the $V$ magnitudes of the stars. The 
agreement between our results and values collected from the literature is 
generally good.
\begin{table}[!t]
\caption{Fitted parameters of the three WDs with positive detections 
of magnetic fields.} 
\smallskip
\begin{center}
{\small
\begin{tabular}[c]{crlccrc}
\tableline
\noalign{\smallskip}
WD & T$_{\rm eff}$ & $\log g$ & $M$ & $d$(spec) &$t_{\mathrm{cool}}$
   & $B_z$ \\
   & (kK) & & (M$_{\odot}$) & (pc) & (Myr) & (G)\\
\noalign{\smallskip}
\tableline
\noalign{\smallskip}
0446$-$789 & 23.45 & 7.72 & 0.49 & 48.8 &   21 &  ~~4283$\pm$640 \\
1105$-$048 & 15.28 & 7.83 & 0.52 & 24.5 &  142 & $-$2134$\pm$447 \\
2359$-$434 &  8.66 & 8.56 & 0.95 &  --- & 2200 & $-$3138$\pm$422 \\
\noalign{\smallskip}
\tableline
\end{tabular}
}
\end{center}
\end{table}

\section{Discussion and Conclusions}
Three of the stars out of 12 normal DA WDs of our sample, WD\,0446-789, 
WD\,1105-048 and WD\,2359-434, exhibit magnetic fields of a few kilogauss in 
one or all available observations. The detection rate of 25~\% suggests 
strongly that a substantial fraction of WDs have a weak magnetic field. 
 
With the exception of the bright WD 40 Eri B, for which a magnetic field of 
only 4\,kG had been detected (Fabrika et al. 2003), our three detections 
have the weakest magnetic fields discovered so far in white dwarfs. Our 
investigation is based on the averaged longitudinal component of the magnetic 
field, meaning that the maximum magnetic field at the WD surface can be 
stronger, depending on the field geometry and on the orientation relative to 
the observer. Therefore, our results for the three objects with a positive 
detection are lower limits, since cancellation effects are expected.

Liebert et al. (2003) have found that the incidence of magnetism at the level 
of $\sim$ 2\,MG or greater is at least $\sim$10\%, or higher. They suggest 
that the total fraction of magnetic WDs may be substantially higher than 10\% 
due to the limited spectropolarimetric analyses capable of detecting lower 
field strengths down to $\sim$ 10\,kG. Our 3 detections out of 12 objects 
seem to indicate that low magnetic fields on WDs ($<$10\,kG) are frequent 
while high magnetic fields are relatively rare. However, with only three 
detections this hypothesis remains insecure. If confirmed by future 
observations, the investigation of weak magnetic fields in WDs could form a 
cornerstone for the future investigation of the properties and evolution of 
stellar magnetic fields.

Our sample of WDs is too small to discuss in detail the dependence of the 
magnetic field strength on the stellar parameters (masses and cooling ages). 
However, two of our detections (WD\,0446$-$789 and WD\,1105$-$048) have 
masses of 0.5 M$_{\odot}$. This means that their progenitors on the 
main-sequence had less than 1 M$_{\odot}$ (Weidemann 2000). These two stars 
are therefore very different from the majority of WDs with MG magnetic fields 
which tend to have higher masses (Greenstein \& Oke 1982; Liebert 1988) and, 
therefore, high-mass parent stars. 

At our signal-to-noise ratio, magnetic fields down to about 2 \,kG can be 
measured. There is still the possibility that all magnetic WDs contain 
surface magnetic fields at the 1\,kG level. To test this hypothesis, much 
longer exposure times would be necessary, even with the VLT.
Full details of our analysis can be found in Aznar Cuadrado et al. (2004).


\end{document}